\documentstyle[11pt]{article}

\newfont{\tenmsb}{msbm10 scaled\magstep1}

\let\ssection=\section\renewcommand{\section}{\setcounter{equation}{0}\ssection}

\newcommand{\half}{{\scriptstyle{\frac{1}{2}}}}

\newcommand{\red}{{\rm red}}
\newcommand{\IC}{{\bf C}}
\newcommand{\const}{\mathop{\rm const}\nolimits}
\newcommand{\parag}{\hfil\break} %%%%% paragraph
\newcommand{\kikezd}{\parag\underbar}

\newcommand{\SO}[1]{{\mathop{\rm SO}}({#1})}

\def\smallover#1/#2{\hbox{$\textstyle{#1\over#2}$}}
\def\p{{\partial}}
\def\vb{{\vec b}}
\def\vc{{\vec c}}

\def\vE{{\vec E}}

\def\vp{{\vec p}}
\def\vP{{\vec P}}
\def\vQ{{\vec Q}}

\def\vx{{\vec x}}
\def\vA{{\vec A}}

\begin{document}

\setlength{\baselineskip}{16pt}

\title{
Exotic galilean symmetry in the non-commutative plane,\\
and the  Hall effect
}

\author{
C.~Duval
\\
Centre de Physique Th\'eorique, CNRS\\
Luminy, Case 907\\
F-13 288 MARSEILLE Cedex 9 (France)
\\ and\\
P.~A.~Horv\'athy
\\
Laboratoire de Math\'ematiques et de Physique Th\'eorique\\
Universit\'e de Tours\\
Parc de Grandmont\\
F-37 200 TOURS (France)
}

\date{\today}

\maketitle

\begin{abstract}
Quantum Mechanics in the non-commutative plane	is shown to admit the
``exotic'' symmetry of the doubly-centrally-extended Galilei group.
When coupled to a planar magnetic field whose strength is the inverse of
the non-commutative parameter, the system becomes singular, and
``Faddeev-Jackiw''
reduction yields the ``Chern-Simons'' mechanics of Dunne, Jackiw, and
Trugenberger.	The reduced  system  moves according to the Hall law.
\end{abstract}

\centerline{
 \texttt{hep-th/0106089} (Revised version)}

\goodbreak

%%%%%%%%%%%%%%%%%%%%%%%%%%%%%%%%%%%%%%%%%%%%%%%%%%%%%%%%%%%%%%%%%%%%%%%%%%%%%%
%%%%%%%%%%%%%%%%%%%%%%%%%%%%%%%%%%%%%%%%%%%%%%%%%%%%%%%%%%%%%%%%%%%%%%%%%%%%%%
\section{Introduction}
%%%%%%%%%%%%%%%%%%%%%%%%%%%%%%%%%%%%%%%%%%%%%%%%%%%%%%%%%%%%%%%%%%%%%%%%%%%%%%
%%%%%%%%%%%%%%%%%%%%%%%%%%%%%%%%%%%%%%%%%%%%%%%%%%%%%%%%%%%%%%%%%%%%%%%%%%%%%%

Quantum Mechanics in the non-commu\-ta\-tive plane has been at the center of
recent interest \cite{noncommQM}. Some formul{\ae} in \cite{NaPo} are, in
particular, rather similar to those we found in \cite{DH}, where we started
with
the two-fold central extension of the planar Galilei group \cite{LL, doubleGal,
LSZ}, labeled by the mass, $m$, and the ``exotic'' parameter, $\kappa$. Then we
argued that a non-relativistic particle in the plane associated to this
``exotic''
Galilei group is endowed with an unconventional structure. Below we point out
that Quantum Mechanics in the non-commutative plane  actually admits our
``exotic'' galilean symmetry; the two models are in fact equivalent, the
non-commutative parameter~$\theta$ being related to the ``exotic'' one
according
to
\begin{equation}
\theta=\frac{\kappa}{m^2}.
\label{exoticnoncomm}
\end{equation}

Coupling an ``exotic'' particle
to an electromagnetic field,  the two extension
parameters combine with the magnetic field, $B$,
into an effective mass, $m^*$, given by (\ref{effmass});
when this latter vanishes,
 we found, furthermore, that
the consistency of the equations of motion requires
that the particle obey the Hall law \cite{DH, QHE}.
Below, we rederive and generalize these results using the
framework of Faddeev and Jackiw \cite{FaJa}.
For $m^*=0$, we get the
 ``Chern-Simons mechanics'' considered some time ago by
 Dunne, Jackiw and Trugenberger \cite{DJT}.

The reduced theory admits the infinite symmetry of area-preserving
diffeomorphisms, found before for the edge currents of the Quantum Hall
states \cite{Winfty}. Finally, we illustrate the general theory on examples.

\goodbreak

%%%%%%%%%%%%%%%%%%%%%%%%%%%%%%%%%%%%%%%%%%%%%%%%%%%%%%%%%%%%%%%%%%%%%%%%%%%%%%
%%%%%%%%%%%%%%%%%%%%%%%%%%%%%%%%%%%%%%%%%%%%%%%%%%%%%%%%%%%%%%%%%%%%%%%%%%%%%%
\section{Exotic symmetry}
%%%%%%%%%%%%%%%%%%%%%%%%%%%%%%%%%%%%%%%%%%%%%%%%%%%%%%%%%%%%%%%%%%%%%%%%%%%%%%
%%%%%%%%%%%%%%%%%%%%%%%%%%%%%%%%%%%%%%%%%%%%%%%%%%%%%%%%%%%%%%%%%%%%%%%%%%%%%%

The fundamental commutation relations for the non-commutative plane
\cite{noncommQM, NaPo} are given by

\begin{equation}
\begin{array}{lll}
	\{x_{1},x_{2}\}=\theta,
	\\
	\{x_{i},p_{j}\}=\delta_{ij},
	\\
	\{p_{1},p_{2}\}=0,
\end{array}
\label{NaPocommrel}
\end{equation}
where $\theta$ is the non-commutative parameter.
The Poisson bracket on phase space,
\begin{equation}
\{f,g\}=
\frac{\p f}{\p\vx}\cdot\frac{\p g}{\p\vp}
-
\frac{\p g}{\p\vx}\cdot\frac{\p f}{\p\vp}
+
\theta\left(
\frac{\p f}{\p x_1}\frac{\p g}{\p x_2}
-
\frac{\p g}{\p x_1}\frac{\p f}{\p x_2}\right),
\label{freeexoticPB}
\end{equation}
differs hence from the canonical one by an additional term. Hamilton's
equations
of a free particle,
$
\dot{\xi}_\alpha=\{\xi_\alpha, h_{0}\}
$
where $h_{0}=\vp{\,}^2/(2m)$ and $\xi=(p_{1},p_{2},x_{1},x_{2})$,
describe therefore
the usual free motion.
Owing to the extra term in (\ref{freeexoticPB}),
some of the conserved quantities
contain additional terms.
The modified angular momentum and Galilean boosts,
\begin{equation}
\begin{array}{ll}
\jmath&=
\displaystyle
\vx\times\vp+\frac{1}{2}\theta\vp{\,}^2+s,
\\[8pt]
g_{i}&=mx_i-p_it+m\theta\,\varepsilon_{ij\,}p_{j}
\end{array}
\label{momentmap}
\end{equation}
(where $s$ is anyonic spin) commute indeed with the free  hamiltonian $h_{0}$.
The key point is that these quantities
satisfy, with the momenta
and the energy, $p_i$ and $h_{0}$,
the ``{\it exotic}'' commutation relations of the
doubly-extended Galilei group \cite{LL, doubleGal},
 which only differ from the standard Galilean commutation
relations in that the boosts close on the exotic parameter
(\ref{exoticnoncomm}) according to
\begin{equation}
\{g_{1},g_{2}\}=-m^2\theta.
\label{boostcr}
\end{equation}

The Hamiltonian framework
presented here is consistent with the acceleration-dependent
Lagrangian
of Lukierski et al. \cite{LSZ}.
This latter is conveniently presented, as
\begin{equation}
L_{0}=\vp\cdot\dot{\vx}-\frac{\vp{\,}^2}{2m}
+\frac{\theta}{2}\,\vp\times\dot{\vp}.
\label{LSZlag}
\end{equation}
Then it is straightforward to show that under a Galilean boost,
$
\vx\to\vx+\vb t,
\
\vp\to\vp+m\vb
$,
the Lagrangian $L_0$ merely changes by a total time derivative,
\begin{equation}
L_0\to L_0+m\frac{\ d}{dt}
\left(\vx\cdot\vb+\frac{1}{2}\vb{\,}^2t+
\frac{\theta}{2}\vb\times\vp\right).
\label{freeexoticchange}
\end{equation}
confirming that the model is indeed non-relativistic.
The model based on the Lagrangian (\ref{LSZlag}) is indeed
equivalent to that constructed in \cite{LL, doubleGal, DH}.

The  conserved quantities are readily recovered by
N{\oe}ther's theorem~: if an infinitesimal transformation
%$\delta\xi, \delta t$
 changes the Lagrangian by a total
time derivative, $\delta L={dC}/{dt}$
(i. e.,  is  a symmetry),
then $({\p L}/{\p\dot{\xi}_{\alpha}})\delta\xi_{\alpha}-h\delta t-C$
is conserved. A rotation leaves $L_{0}$
 invariant, so that the exotic contibution
to the angular momentum in (\ref{momentmap}) comes from the
$({\p L}/{\p\dot{\xi}_{\alpha}})\delta\xi_{\alpha}$ term alone.
For a boost,
half of the exotic contibution comes from the latter term,
and the other
half from the response (\ref{freeexoticchange}) of the Lagrangian.

The doubly-centrally-extended---or ``exotic''---Galilei group can be
conveniently represented by the group of the $6\times6$ matrices
\begin{equation}
a=\left(
\begin{array}{ccccc}
\displaystyle
A&\vb&0&\vc&\half\varepsilon\vb\\[8pt]
\displaystyle
0&1&0&e&0\\[8pt]
\displaystyle
\vb\cdot A&\half\vb{\,}^2&1&u&v\\[8pt]
0&0&0&1&0\\[8pt]
0&0&0&0&1
\end{array}
\right),
\label{BargmannMatrices}
\end{equation}
where $\varepsilon$ is the matrix
$\big(\varepsilon_{ij}\big)$;
$A\in\SO{2}$
represents a planar rotation, $\vb$ a Galilean boost, $\vc$ a
space translation, and $e$ a time translation;
$u$ and $v$ parametrize the two-dimensional center.
The classical phase space with Poisson structure (\ref{freeexoticPB}) can be
identified, as in \cite{doubleGal, DH}, with a coadjoint orbit of the
doubly-extended Galilei group (\ref{BargmannMatrices}) defined by the
invariants
$m$ and $\kappa$, cf.~ (\ref{exoticnoncomm}).

Due to the presence of the ``exotic'' term there is no position representation.
Our clue is to observe that the $p_i$ and
\begin{equation}
Q_i=x_i+\frac{1}{2}\theta\varepsilon_{ij}\,p_j
\label{freecancoord}
\end{equation}
are canonical coordinates so that they satisfy the {\it ordinary}
relations (\ref{NaPocommrel}) with $\theta=0$.
Canonical quantization yields hence, in the momentum picture, that the
quantum operator $\widehat{p_i}$ is multiplication by
$p_i$, and (setting $\hbar=1$):
\begin{equation}
\widehat{x_j}=\widehat{Q_j}
-\displaystyle\frac{1}{2}\theta\varepsilon_{jk}\,\widehat{p_k}
=
i\displaystyle\frac{\p}{\p p_j}
-\displaystyle\frac{1}{2}\theta\varepsilon_{jk}\,p_k.
\end{equation}

Putting
$G_i=g_i-({m\theta}/{2})\varepsilon_{ij}\,p_j$
we get $\big\{G_i,G_j\big\}=0$, while the other commutation relations
remain unchanged: we obtain the ordinary (singly-extended) Galilei
algebra  \cite{doubleGal}.
The Hamiltonian, $\widehat{h_0}=\vp{\,}^2/(2m)$, is hence standard.
Unlike its classical counterpart in (\ref{momentmap}),
the quantum angular momentum retains the usual form
$
\widehat{\jmath}=-i\varepsilon_{jk}\,p_{j}
\p_{p_k}+s
$,
whereas the ``exotic'' contribution only appears in the boosts,
namely
\begin{equation}
	\widehat{g_j}	=
	m\left[i\displaystyle\frac{\ \p }{\p p_{j}}
	+\displaystyle\frac{1}{2}\theta\varepsilon_{jk}\,p_{k}\right].
\label{exoboost}
\end{equation}
 The factor  $1/2$ here w.r.t. the classical expression (\ref{momentmap})
is explained by
$\hat{g}_{i}=m\hat{x}_{i}+m\theta\varepsilon_{ij}p_{j}
=
m\hat{Q}_{i}+\half m\theta\varepsilon_{ij}p_{j}$.
Completed with the mass, $m$, and the ``exotic'' parameter,
$\kappa=m^2\theta$, these operators
span the  ``exotic'' Galilei algebra \cite{LL, doubleGal}.

The associated irreducible unitary representation $U_{m,\theta}$ of the matrix
group (\ref{BargmannMatrices}), on the space of wave functions~$\psi(\vp)$,
is deduced accordingly,
\begin{equation}
U_{m,\theta}(a)\psi(\vp)=
\exp\left(
i\left[\frac{\vp{\,}^2e}{2m}-\vp\cdot\vc+s\varphi+m\,u\right]
+
im\theta\left[\frac{1}{2}\vb\times\vp+m\,v\right]
\right)
\psi\left(A^{-1}(\vp-m\vb\,)\right),
\label{exoticrep}
\end{equation}
where $\varphi$ is the angle of the rotation $A$, see also~\cite{doubleGal}.

\goodbreak

%%%%%%%%%%%%%%%%%%%%%%%%%%%%%%%%%%%%%%%%%%%%%%%%%%%%%%%%%%%%%%%%%%%%%%%%%%%%%%
%%%%%%%%%%%%%%%%%%%%%%%%%%%%%%%%%%%%%%%%%%%%%%%%%%%%%%%%%%%%%%%%%%%%%%%%%%%%%%
\section{Coupling to a gauge field}
%%%%%%%%%%%%%%%%%%%%%%%%%%%%%%%%%%%%%%%%%%%%%%%%%%%%%%%%%%%%%%%%%%%%%%%%%%%%%%
%%%%%%%%%%%%%%%%%%%%%%%%%%%%%%%%%%%%%%%%%%%%%%%%%%%%%%%%%%%%%%%%%%%%%%%%%%%%%%

As found in \cite{DH} using Souriau's symplectic framework
\cite{SSD}, minimal coupling to an electro\-magnetic field $(\vE,B)$
unveils new and surprising features. Here we explain this
using the ``Faddeev-Jackiw''  formalism \cite{FaJa}.
Let us hence generalize the free expression (\ref{LSZlag})
by considering the action
\begin{equation}
 \int{
(\vp-\vA\,)\cdot d\vx
-h\,dt
+
\frac{\theta}{2}\,\vp\times d\vp
},
\label{matteraction}
\end{equation}
where $(V,\vA)$ is an electro-magnetic
potential, the Hamiltonian being given by
\begin{equation}
h=\frac{\vp{\,}^2}{2m}+V.
\label{Hamiltonian}
\end{equation}

 The associated Euler-Lagrange equations read
\begin{equation}
\left\{\!
\begin{array}{rcl}\displaystyle
m^*\dot{x}_{i}
&=&
p_{i}-\displaystyle m\theta\,\varepsilon_{ij}E_{j},
\\[8pt]
\displaystyle
\dot{p}_{i}
&=&
E_{i}+B\,\varepsilon_{ij}\dot{x}_{j},
\end{array}
\right.
\label{eqmotion}
\end{equation}
where we have introduced the \textit{effective mass}
\begin{equation}
m^*=m(1-\theta B).
\label{effmass}
\end{equation}
The velocity and momentum
are different if $\theta\neq0$.
The equations of motions (\ref{eqmotion}) can also be written as
\begin{equation}
	\omega_{\alpha\beta}\dot{\xi}_\beta=\frac{\p h}{\p \xi_\alpha},
\qquad\hbox{where}\qquad
	\big(\omega_{\alpha\beta}\big)=
	\left(\begin{array}{cccc}
	0&\theta&1&0\\[2mm]
	-\theta&0&0&1\\[2mm]
	-1&0&0&B\\[2mm]
	0&-1&-B&0\\[2mm]
	\end{array}\right).
\label{symplecticmatrix}
\end{equation}
Note that the electric and magnetic fields
are otherwise arbitrary solutions of the homogeneous Maxwell equation
$\p_tB+\varepsilon_{ij}\p_iE_j=0$, which guarantees that the
two-form $\omega=\half\omega_{\alpha\beta}d\xi^\alpha\wedge{}d\xi^\beta$
is closed, $d\omega=0$.
Our matrix (\ref{symplecticmatrix}) is in fact
$(m^*/m)$--times that posited in \cite{NaPo}.

\goodbreak

%%%%%%%%%%%%%%%%%%%%%%%%%%%%%%%%%%%%%%%%%%%%%%%%%%%%%%%%%%%%%%%%%%%%%%%%%%%%%%
%\subsection{Nonzero effective mass}
%%%%%%%%%%%%%%%%%%%%%%%%%%%%%%%%%%%%%%%%%%%%%%%%%%%%%%%%%%%%%%%%%%%%%%%%%%%%%%

When  $m^*\neq0$, the determinant
  $\det\big(\omega_{\alpha\beta}\big)=
\left(1-\theta\, B\right)^2
=\big(m^*/m\big)^2$
is nonzero;
 the matrix  $(\omega_{\alpha\beta})$ in (\ref{symplecticmatrix})
 is indeed symplectic, and can therefore be inverted.
 Then
the equations of motion
(\ref{symplecticmatrix}) (or (\ref{eqmotion}))
take the form $\dot{\xi}_{\alpha}=\big\{\xi_{\alpha}, h\big\}$,
with the standard Hamiltonian (\ref{Hamiltonian}), but with the new Poisson
bracket $\{f,g\}=(\omega^{-1})_{\alpha\beta}\p_{\alpha}f\p_{\beta}g$ which
reads,
explicitly,
\begin{equation}
\begin{array}{rcl}
\{f,g\}
&=&
\displaystyle\frac{m}{m^*}\left[
\frac{\p f}{\p\vx}\cdot\frac{\p g}{\p\vp}
-
\frac{\p g}{\p\vx}\cdot\frac{\p f}{\p\vp}
\right.\\[12pt]
&&
+
\left.
{\theta}\left(
\displaystyle\frac{\p f}{\p x_1}\frac{\p g}{\p x_2}
-
\frac{\p g}{\p x_1}\frac{\p f}{\p x_2}\right)
+
B\displaystyle\left(
\frac{\p f}{\p p_1}\frac{\p g}{\p p_2}
-
\frac{\p g}{\p p_1}\frac{\p f}{\p p_2}\right)
\right].\hfill
\end{array}
\label{exoticPB}
\end{equation}
Note that the fundamental commutation relations (\ref{NaPocommrel})
are now modified \cite{DH} as
 \begin{equation}
\begin{array}{lll}
\{x_{1},x_{2}\}=
\displaystyle\frac{m}{m^*}\,	\theta,
	\\[3.4mm]
	\{x_{i},p_{j}\}=\displaystyle\frac{m}{m^*}\,\delta_{ij},
	\\[3.4mm]
	\{p_{1},p_{2}\}=\displaystyle\frac{m}{m^*}\,B.
\end{array}
\label{Bcommrel}
\end{equation}
Thus, both the coordinates and the momenta span independent Heisenberg
algebras.
\goodbreak

Further insight can be gained when the magnetic field $B$
is a (positive) nonzero constant, which turns out the most interesting case,
and will be henceforth assumed.
The vector potential can then be chosen
as $A_i=\half{}B\varepsilon_{ij}\,x_{j}$, the electric field $E_i=-\p_iV$ being
still arbitrary. Introducing the new coordinates somewhat similar to those in
(\ref{freecancoord}),
\begin{equation}
\left\{\begin{array}{c}
Q_{i}=x_{i}+
\displaystyle\frac{1}{B}\left[1-\sqrt{
\displaystyle\frac{m^*}{m}}\,\right]
\varepsilon_{ij}\,p_{j},\\[8pt]
P_{i}=\sqrt{\displaystyle\frac{m^*}{m}}\,p_i-
\displaystyle\frac{1}{2}B\varepsilon_{ij}\,Q_{j},\hfill
\end{array}\right.
\label{goodcoordinates}
\end{equation}
will allow us to generalize our results in
\cite{DH} from a constant to any electric field.

Firstly, the ``Cartan'' one-form \cite{SSD} in the action (\ref{matteraction})
reads simply
$
P_idQ_i-hdt,
$
so that the symplectic form on phase space retains the canonical guise,
$\omega=dP_{i}\wedge dQ_{i}$.
The price to pay is that the
Hamiltonian becomes rather complicated,
\begin{equation}
h=
\frac{1}{2m^*}\left(\vec{P}+\half{}B\varepsilon\vec{Q}\,\right)^2
+
V\left(\alpha\vec{Q}+\beta\varepsilon\vec{P}\,\right),
\label{bigham}
\end{equation}
with
$ \alpha=\frac{1}{2}(1+\sqrt{{m}/{m^*}})
$
and
$
\beta=B^{-1}(1-\sqrt{{m}/{m^*}}).
$

The equations of motion (\ref{eqmotion}) are conveniently presented
in terms of the new variables $\vQ$ and the old momenta $\vp$, as
\begin{equation}
    \left\{\begin{array}{ll}
\dot{Q}_{i}=\varepsilon_{ij}\displaystyle\frac{E_{j}}{B}
+\displaystyle\sqrt{\frac{m}{m^*}}\left(
\frac{p_{i}}{ m}-\varepsilon_{ij}\displaystyle\frac{E_{j}}{B}\right),
\\[4mm]
\dot{p}_{i}=
\varepsilon_{ij}{B}\displaystyle\frac{m}{m^*}\left(
\frac{p_{j}}{ m}-
\varepsilon_{jk}\displaystyle\frac{E_{k}}{B}\right).
\end{array}\right.
\label{Qpeqmot}
\end{equation}
Note that all these expressions diverge when
$m^*$ tends to zero.

%%%%%%%%%%%%%%%%%%%%%%%%%%%%%%%%%%%%%%%%%%%%%%%%%%%%%%%%%%%%%%%%%%%%%%%%%%%%%%
%\subsection{Vanishing effective mass: Hamiltonian reduction}
%%%%%%%%%%%%%%%%%%%%%%%%%%%%%%%%%%%%%%%%%%%%%%%%%%%%%%%%%%%%%%%%%%%%%%%%%%%%%%

When the magnetic field takes the particular value
\begin{equation}
B=B_{c}=\frac{1}{\theta},
\label{critB}
\end{equation}
the effective mass (\ref{effmass}) vanishes, $m^*=0$, so that
$\det(\omega_{\alpha\beta})=0$, and the system becomes singular.
Then the time derivatives $\dot{\xi_\alpha}$ can no longer be
expressed from the
variational equations (\ref{symplecticmatrix}), and we have resort to
``Faddeev-Jackiw'' reduction \cite{FaJa}. In accordance with the Darboux
theorem (see, e.g.,~\cite{SSD}), the Cartan one-form
in (\ref{matteraction})
can be written, up to an exact term, as $\vartheta-hdt$,
with
$
\vartheta=
(p_{i}-\half{B_{c}}\,\varepsilon_{ij}\,x_{j})dx_{i}
+\frac{1}{2}\theta\varepsilon_{ij}\,p_{i}dp_{j}
=
P_{i}dQ_{i},
$
where the new coordinates read, consistently with
(\ref{goodcoordinates}),
\begin{equation}
Q_{i}=x_{i}+\displaystyle\frac{1}{B_{c}}\varepsilon_{ij}p_{j},
\label{redcancoord}
\end{equation}
while the
$P_{i}=-\half{}B_{c}\,\varepsilon_{ij}\,Q_{j}$
are in fact the rotated coordinates $Q_{i}$.
Eliminating the original coordinates $\vx$ using (\ref{redcancoord}),
we see that the Cartan one-form reads
$
P_idQ_i-H(\vQ,\vp)dt,
$
where
$
H(\vQ,\vp)=\vp{\,}^2/(2m)+V(\vQ,\vp).
$
As the $p_{i}$ appear here with no derivatives,
they can be eliminated using their equation of motion
$
\p H(\vQ,\vp)/\p\vp=0,
$
namely
\begin{equation}
\frac{p_i}{m}-\frac{\varepsilon_{ij}E_j}{B_{c}}=0,
\label{pHall}
\end{equation}
cf. (\ref{Qpeqmot}).
 Inserting (\ref{pHall}) into (\ref{redcancoord}) and taking partial
derivatives,
we find
\begin{eqnarray*}
\frac{\p Q_j}{\p x_i}=\delta_{ji}-\frac{m}{B_{c}^2}\frac{\p E_j}{\p x_i},
\qquad
\frac{\p H}{\p x_i}=
m\frac{E_j}{B_{c}}\frac{\p E_j}{\p x_i}-E_i.
\end{eqnarray*}
Hence
${\p H}/{\p\vQ}=(\p H/\p\vx)\cdot(\p\vx/\p\vQ)=-\vE$.
Consequently, the reduced
Hamiltonian is (modulo a constant) just the original potential,
viewed as a function of the ``twisted'' coordinates $\vQ$, viz
\begin{equation}
H=V(\vQ).
\label{redham}
\end{equation}
This rule is referred to as the ``Peierls substitution''
\cite{DJT, DH}.
Since $\p^2H/\p{}p_i\p{}p_j=\delta_{ij}/m$ is
already non singular,
the reduction stops, and we end up with the reduced Lagrangian
\begin{equation}
L_{\rm red}=\frac{1}{2\theta}\vQ\times\dot{\vQ}-V(\vQ),
\label{redlag}
\end{equation}
supplemented with the Hall constraint (\ref{pHall}).
The $4$-dimensional phase space is hence reduced to $2$ dimensions, with
$Q_1$ and $Q_2$ in (\ref{redcancoord})
as canonical coordinates, and reduced symplectic two-form
$
\omega_\red=
\half B_{c}\,\varepsilon_{ij}dQ_{i}\wedge dQ_{j}
$
so that the reduced Poisson bracket is
\begin{equation}
\big\{F, G\big\}_{\red}=-\frac{1}{B_{c}}
\Big(\frac{\p F}{\p Q_{1}}\,\frac{\p G}{\p Q_{2}}
-\frac{\p G}{\p Q_{1}}\,\frac{\p F}{\p Q_{2}}\Big).
\label{redpoisson}
\end{equation}
The twisted coordinates are therefore again
non-commuting,
\begin{equation}
\big\{Q_{1}, Q_{2}\big\}_{\red}=-\theta=-\frac{1}{B_{c}},
\label{redcommrel}
\end{equation}
cf. (\ref{boostcr}).
The equations of motion associated with (\ref{redlag}), and also consistent
with
the Hamilton equations $\dot{Q}_i=\big\{Q_i,H\}_{\red}$, are given by
\begin{equation}
\dot{Q}_{i}=\varepsilon_{ij}\frac{E_{j}}{B_{c}},
\label{QHall}
\end{equation}
in accordance with the Hall law (compare (\ref{Qpeqmot}) with the
divergent terms removed).

\goodbreak

Putting $B_{c}=1/\theta$, the Lagrangian (\ref{redlag}) becomes formally
identical to the one
Dunne et al.~\cite{DJT} derived letting the {\it real} mass
go to zero. Note, however, that while $\vQ$ denotes
real position in Ref.~\cite{DJT}, our $\vQ$ here is  the
``twisted'' expression  (\ref{redcancoord}), with the magnetic field
frozen at the critical value $B_{c}=1/\theta$, determined by the ``exotic''
structure.
\goodbreak

%%%%%%%%%%%%%%%%%%%%%%%%%%%%%%%%%%%%%%%%%%%%%%%%%%%%%%%%%%%%%%%%%%%%%%%%%%%%%%
%%%%%%%%%%%%%%%%%%%%%%%%%%%%%%%%%%%%%%%%%%%%%%%%%%%%%%%%%%%%%%%%%%%%%%%%%%%%%%
\section{Quantization}
%%%%%%%%%%%%%%%%%%%%%%%%%%%%%%%%%%%%%%%%%%%%%%%%%%%%%%%%%%%%%%%%%%%%%%%%%%%%%%
%%%%%%%%%%%%%%%%%%%%%%%%%%%%%%%%%%%%%%%%%%%%%%%%%%%%%%%%%%%%%%%%%%%%%%%%%%%%%%

Let us conclude our general theory by the quantization of the coupled system.
Again, owing to the exotic term, the position representation does not exist.
We can use, instead, the twisted coordinates $\vQ$ in (\ref{goodcoordinates});
and consider wave functions as simply depending on~$\vQ$. Quantizing the
Hamiltonian (\ref{bigham}) is, however a rather tough task: apart from
the ``gentle'' quadratic kinetic term, one also
has to quantize the otherwise arbitrary function
$V(\alpha\vQ+\beta{}\varepsilon\vP\,)$ of the conjugate
variables
$\vP$ and $\vQ$. This goes beyond our scope here;
we focus, therefore, our attention to the kinetic term.

%%%%%%%%%%%%%%%%%%%%%%%%%%%%%%%%%%%%%%%%%%%%%%%%%%%%%%%%%%%%%%%%%%%%%%%%%%%%%%
%\subsection{Projection to LLL: $m^*\to0$}
%%%%%%%%%%%%%%%%%%%%%%%%%%%%%%%%%%%%%%%%%%%%%%%%%%%%%%%%%%%%%%%%%%%%%%%%%%%%%%

Introducing the complex coordinates
\begin{equation}
\left\{\begin{array}{c}
z=\displaystyle\frac{\sqrt{B}}{2}\big(Q_1+iQ_2\big)
+\displaystyle\frac{1}{\sqrt{B}}\big(-iP_1+P_2)\hfill
\\[3mm]
w=\displaystyle\frac{\sqrt{B}}{2}\big(Q_1-iQ_2\big)
+\displaystyle\frac{1}{\sqrt{B}}\big(-iP_1-P_2)\hfill
\end{array}\right.
\label{bigcomplexcoord}
\end{equation}
the two-form $dP_i\wedge dQ_i$ on $4$-dimensional phase space becomes the
canonical K\"ahler two-form of $\IC^2$, viz
$
\omega=(2i)^{-1}\big(d\bar{z}\wedge dz+d\bar{w}\wedge dw\big).
$
 Then geometric quantization \cite{SSD, GQ} yields,
with the choice of the antiholomorphic polarization, the ``unreduced''
quantum Hilbert space, consisting of
the ``Bargmann-Fock'' wave functions
\begin{equation}
	\psi(z,\bar{z},w,\bar{w})
	=f(z,w)e^{-\frac{1}{4}(z\bar{z}+w\bar{w})},
\label{bigBF}
\end{equation}
where $f$ is holomorphic in both of its variables.
The fundamental quantum operators,
\begin{equation}
	\left\{\begin{array}{ll}
		\widehat{z}\,f=zf,
		&\widehat{\bar{z}}\,f=2\partial_{z}f,
		\\[8pt]
		\widehat{w}f=wf,
		&\widehat{\!\bar{w}}f=2\partial_{w}f,
	\end{array}\right.
	\label{bigmomop}
\end{equation}
satisfy the commutation relations
$\big[\widehat{\bar{z}},\widehat{z}\big]
=\big[\,\widehat{\!\bar{w}},\widehat{w}\big]
=2$,
and
$
\big[\widehat{z},\widehat{w}\big]
=\big[\widehat{\bar{z}},\widehat{\!\bar{w}}\big]
=0
$.
 We recognize here the familiar creation and annihilation operators, namely
$a^*_z=z$, $a^*_w=w$, and $a_z=\p_z$, $a_w=\p_w$.

Using (\ref{goodcoordinates}), the (complex)  momentum
$p=p_{1}\!+\! ip_{2}$
and the kinetic part, $h_0$, of the Hamiltonian~(\ref{Hamiltonian}) become,
respectively,
\begin{equation}
	p=-i\sqrt{\frac{mB}{m^*}}\,\bar{w}
	\qquad\hbox{and}\qquad
	h_{0}=
	\frac{B}{2m^*}\,w\bar{w}.
\label{complexmomentum}
\end{equation}

For $m^*\neq0$ the wave function satisfies the Schr\"odinger
equation
$
i\p_tf=\widehat{h}f,
$
with
$
\widehat{h}=\widehat{h}_{0}+\widehat{V}.
$
The quadratic kinetic term here is
 \begin{equation}
	 \widehat{h}_{0}=
	 \frac{B}{4m^*}\big(\widehat{w}\,\widehat{\!\bar{w}}+
	 \widehat{\!\bar{w}}\,\widehat{w}\big)=
	 \frac{B}{2m^*}\big(\widehat{w}\,\widehat{\!\bar{w}}+1\big).
\label{kinham}
\end{equation}

The case when the effective mass tends to zero is conveniently studied in this
framework. On the one hand, in the limit $m^*\to0$, one has
\begin{equation}
	z\to\sqrt{B}\,Q,
	\qquad
	w\to 0,
\end{equation}
where $Q=Q_{1}+iQ_{2}$, cf. (\ref{goodcoordinates}); the $4$-dimensional
phase space reduces to the complex plane.
On the other hand, from
(\ref{complexmomentum}) and (\ref{bigmomop}) we deduce that
\begin{equation}
	i\sqrt{\frac{m^*}{mB}}\,\widehat{p}=\widehat{\!\bar{w}}=2\p_{w}.
\end{equation}
The limit $m^*\to0$ is hence enforced, at the quantum level,
by requiring that the wave functions be independent of the
coordinate $w$, i.e.,
\begin{equation}
\p_{w}f=0,
\label{LLLcond}
\end{equation}
yielding the reduced wave functions of the form
\begin{equation}
\Psi(z,\bar{z})=f(z)e^{-\frac{1}{4}z\bar{z}},
\label{redwavefunction}
\end{equation}
where $f$ is a holomorphic function of the reduced phase space parametrized by
$z$.
When viewed in the ``big'' Hilbert space (see~(\ref{bigBF})), these wave
functions
belong, by (\ref{kinham}), to the lowest Landau level \cite{QHE, GJ, DH}.

Using the fundamental operators $\widehat{z}$ an $\widehat{\bar{z}}$ given in
(\ref{bigmomop}), we easily see that the (complex) ``physical'' position
$x=x_{1}+ix_{2}$ and its quantum counterpart $\widehat{x}$, namely
\begin{equation}
x=\frac{1}{\sqrt{B_c}}\left(z+\sqrt{\frac{m}{m^*}}\,\bar{w}\right),
	\qquad
\widehat{x}=
\frac{1}{\sqrt{B_c}}\left(z+\sqrt{\frac{m}{m^*}}\,2\partial_{w}\right),
\label{complexposition}
\end{equation}
manifestly diverge when $m^*\to0$.
{\it Positing from the outset} the conditions (\ref{LLLcond})
 the divergence is suppressed, however,
 leaving us with the reduced position operators
\begin{equation}
	\widehat{x}\,f=\widehat{Q}f=\frac{1}{\sqrt{B_c}}\,zf,
\qquad
	\widehat{\bar{x}}\,f=\widehat{\!\bar{Q}}f=
	\frac{2}{\sqrt{B_c}}\,\p_zf,
\label{redquantumposition}
\end{equation}
whose commutator is
$[\widehat{Q},\widehat{\!\bar{Q}}]={2}/{B_c},$
cf. (\ref{redcommrel}).
In conclusion, we recover the
``Laughlin'' description \cite{QHE}
of the ground states of the FQHE.
(In \cite{DH}, these  results have been
obtained by quantizing the reduced model.)
Quantization of the reduced Hamiltonian (which is, indeed, the
potential $V(z,\bar{z})$),
can be achieved using, for instance, anti-normal ordering  \cite{QHE,
GJ, DJT}.
\goodbreak

%%%%%%%%%%%%%%%%%%%%%%%%%%%%%%%%%%%%%%%%%%%%%%%%%%%%%%%%%%%%%%%%%%%%%%%%%%%%%%
%%%%%%%%%%%%%%%%%%%%%%%%%%%%%%%%%%%%%%%%%%%%%%%%%%%%%%%%%%%%%%%%%%%%%%%%%%%%%%
\section{Examples}
%%%%%%%%%%%%%%%%%%%%%%%%%%%%%%%%%%%%%%%%%%%%%%%%%%%%%%%%%%%%%%%%%%%%%%%%%%%%%%
%%%%%%%%%%%%%%%%%%%%%%%%%%%%%%%%%%%%%%%%%%%%%%%%%%%%%%%%%%%%%%%%%%%%%%%%%%%%%%

%%%%%%%%%%%%%%%%%%%%%%%%%%%%%%%%%%%%%%%%%%%%%%%%%%%%%%%%%%%%%%%%%%%%%%%%%%%%%%
%\subsection{Constant electric field}
%%%%%%%%%%%%%%%%%%%%%%%%%%%%%%%%%%%%%%%%%%%%%%%%%%%%%%%%%%%%%%%%%%%%%%%%%%%%%%

In both examples studied below, we will consider a particle with unit mass,
$m=1$, (and unit charge, as before).

The simplest non-trivial example
 is provided by a constant
electric field \cite{DH}.
For nonvanishing effective mass $m^*\neq0$,
the equations of motion (\ref{Qpeqmot}) readily
 imply that the ``position''
\begin{equation}
	R=R_{1}+iR_{2},
	\qquad\hbox{where}\qquad
R_{i}=Q_{i}-\frac{1}{B}\varepsilon_{ij}E_{j}\,t,
\label{Hcoordinates}
\end{equation}
(as well as the momentum, $\vp$) rotates in the plane with frequency
$
B/m^*$, viz
 $R(t)=e^{-i(B/m^*)t}R_{0}.$
In the twisted coordinates
$Q_{i}=R_{i}+\varepsilon_{ij}(E_{j}/B)\,t$, the motion is therefore the usual
cyclotronic motion (with modified frequency),
while the guiding center drifts with the Hall velocity
$\varepsilon_{ij}E_{j}/B$.

When $m^*=0$, the reduced equation (\ref{QHall}) requires simply
$
\dot{R}_{i}=0
$~:
 the rotation is eliminated, and we are left, cf.~(\ref{pHall}), with
the uniform drift of the guiding center alone,
\begin{equation}
x_{i}(t)=
\varepsilon_{ij}\frac{E_{j}}{B}\,t
+
x_i(0).
\end{equation}

The canonical transformations of the reduced Poisson bracket (\ref{redpoisson})
coincide with the symplectic transformations of the plane.
These latter are in fact generated by  the observables, i.e., the (smooth)
functions~$F(R)$ of the variable $R$ in (\ref{Hcoordinates}).  But, owing
to the
particular time dependence of~$R$ in (\ref{Hcoordinates}),
$
\{F, H\}=\p_{t}F
$
for {\it any} function  $F(R)$, which generates, therefore, a symmetry. In the
plane, symplectic and area-preserving transformations coincide, yielding the
$w_\infty$ symmetry \cite{Winfty}.

Examples of observables $F$ linear in $Q$ include
 the reduced energy, $H=-\vec{E}\cdot\vQ$, and
 the reduced momenta, $\Pi_i=B\varepsilon_{ij}Q_{j}+E_{j}t$.
These latters  have the Poisson
bracket of ``magnetic translations'', $\{\Pi_1,\Pi_2\}_\red=B_c$, see
(\ref{Bcommrel}).
The quadratic observables generate, in turn, the well-known
${\rm  sp}(1)$-symmetry of the $1$-dimensional harmonic
oscillator.

It is worth pointing out that
the reduced Hilbert space will be acted upon by
$W_{\infty}$, the quantum version of the classical symmetry
algebra $w_{\infty}$.

\goodbreak

%%%%%%%%%%%%%%%%%%%%%%%%%%%%%%%%%%%%%%%%%%%%%%%%%%%%%%%%%%%%%%%%%%%%%%%%%%%%%%
%\subsection{Exotic Oscillator in a planar magnetic field}
%%%%%%%%%%%%%%%%%%%%%%%%%%%%%%%%%%%%%%%%%%%%%%%%%%%%%%%%%%%%%%%%%%%%%%%%%%%%%%

As another illustration, let us describe an ``exotic'' particle moving in a
constant magnetic field, $B$, and a harmonic potential
$
V(\vx\,)=\half\omega^2\vx{\,}^2,
$
cf. \cite{DJT, NaPo}.
%
%%%%%%%%%%%%%%%%%%%%%%%%%%%%%%%%%%%%%%%%%%%%%%%%%%%%%%%%%%%%%%%%%%%%%%%%%%%%%%
%\subsection{Classical motions}
%%%%%%%%%%%%%%%%%%%%%%%%%%%%%%%%%%%%%%%%%%%%%%%%%%%%%%%%%%%%%%%%%%%%%%%%%%%%%%
%
  Let us first assume that the effective mass does not vanish,
$m^*\neq0$. The equations of motion (\ref{eqmotion}), {\it viz}.
 \begin{equation}
m^*\ddot{x}_{i}=B^*\epsilon_{ij}\dot{x}_{j}-\omega^{2}x_{i},
\qquad
{\rm where}
\qquad
B^*=B+{\theta}\omega^{2},
\label{heqmot}
\end{equation}
 describe  an
ordinary, non ``exotic'', particle
with (effective) mass $m^*$, moving in a combined
``effective magnetic field'' $B^*$
and harmonic field $\vec{E}=-\omega^{2}\vx$. Our particle evolves according to
\begin{equation}
x(t)
=
e^{-i(B^*/2m^*)t}
\left[C\cos\omega^{*}t+D\sin\omega^{*}t\right],
\qquad
{\rm with}
\qquad
\omega^*=
\sqrt{\left(\frac{B^*}{2m^*}\right)^2+\frac{\omega^{2}}{m^*}
},
\label{ellmotion}
\end{equation}
where $C$ and $D$ are complex constants; see also~\cite{DJT}.
The elliptic trajectories described  in
the square bracket are hence combined with a circular motion
represented by the exponential factor.

The system is plainly symmetric with respect to planar rotations;
the conserved angular momentum
(consistent with that in \cite{NaPo}) reads
\begin{equation}
j=\vx\times\vp+\frac{\theta}{2}\vp{\,}^2
+\frac{B}{2}{\vx}{\,}^2+s.
\label{hangmom}
\end{equation}
Note here the term coming from the ``exotic'' structure,
and also the ``spin from isospin''
contribution due to the symmetric magnetic field.
As to quantization,
it is enough to replace $B$ and~$m$ by $B^*$ and $m^*$ in the
formul{\ae} of Dunne et al. \cite{DJT}.

When the magnetic field takes the critical value $B_{c}=1/\theta$, the
effective
mass vanishes and the motion obeys the reduced equation.
The ``twisted'' coordinates in (\ref{redcancoord}) are now
proportional to the original ``physical'' position,
$
\vQ=\big(1+\theta^2\omega^2\big)\vx.
$
Hence, the motion is governed by the same equations,
namely
\begin{equation}
\dot{Q}_{i}=-\omega_{c}^*\,\varepsilon_{ij}\,Q_{j}
\qquad\hbox{where}\qquad
\omega_{c}^*=\frac{\theta\omega^{2}}{1+\theta^2\omega^2}.
\label{Qeq}
\end{equation}
Putting $Q=Q_{1}+iQ_{2}$, we find
\begin{equation}
Q(t)=e^{-i\omega_{c}^* t}Q_{0}
\label{Qsol}
\end{equation}
with $Q_0$ a complex constant. All particles move collectively, namely
 along circles perpendicular to the electric field,
with uniform angular velocity $\omega_{c}^*$.
Intuitively, for $m^*=0$,
the general elliptic trajectories (\ref{ellmotion})
are forbidden, leaving us with
the simple circular  motions only.
The reduced symplectic form and Hamiltonian are, respectively,
\begin{equation}
	\Omega=
\frac{1}{\theta}\,dQ_{1}\wedge dQ_{2},
\qquad\hbox{and}\qquad
H=\frac{\omega^{2}}
{2(1+\theta^2\omega^{2})}\,{\vQ}^2.
\label{redhamsymp}
\end{equation}
Since the Hall constraint (\ref{pHall}) is consistent with rotational
symmetry, the reduced system will have a conserved angular
momentum (which turns out to be proportional to the reduced Hamiltonian).
We get hence, once again, a $1$-dimensional harmonic oscillator
with, this time, the usual quadratic Hamiltonian.
Its spectrum is, therefore \cite{DJT},
\begin{equation}
	E_{n}=\frac{\theta\omega^2}{1+\theta^2\omega^2}\left(n+\frac{1}{2}\right),
\qquad
	n=0,1,\ldots
\end{equation}
At last, the  $w_{\infty}$ symmetry of the reduced model discussed
above  is now generated by the functions $F(Q_0)$, see (\ref{Qsol}).

%%%%%%%%%%%%%%%%%%%%%%%%%%%%%%%%%%%%%%%%%%%%%%%%%%%%%%%%%%%%%%%%%%%%%%%%%%%%%%
%%%%%%%%%%%%%%%%%%%%%%%%%%%%%%%%%%%%%%%%%%%%%%%%%%%%%%%%%%%%%%%%%%%%%%%%%%%%%%
\section{Discussion}
%%%%%%%%%%%%%%%%%%%%%%%%%%%%%%%%%%%%%%%%%%%%%%%%%%%%%%%%%%%%%%%%%%%%%%%%%%%%%%
%%%%%%%%%%%%%%%%%%%%%%%%%%%%%%%%%%%%%%%%%%%%%%%%%%%%%%%%%%%%%%%%%%%%%%%%%%%%%%

Our approach allows us to rederive some previous formulae starting with first
principles. If one accepts that any theory consistent with the (extended)
Galilean symmetry is physical, there is no reason to discard our ``exotic''
Galilean theory. The rather obvious equivalence of ``non-commutative'' and
``exotic'' approaches found here is important, as it allows for a
``technology transfer''.

The interplay between the ``exotic'' and
the magnetic terms in (\ref{exoticPB})
which leads, for $B=B_{c}$, to the singular behavior studied above
happens precisely when
$
\vp\to-{\vx}/{\theta},
\
\vx\to\theta\,\vp
$
is a canonical transformation that merely
interchanges the magnetic and the ``exotic'' terms
in (\ref{symplecticmatrix}).

It is also rather intriguing to observe that the Hall
motions discussed
above are, strictly speaking, not the only possible motions of the system.
 Let us indeed assume that $m^*=0$,
and consider an ``exotic'' particle whose initial velocity is
{\it inconsistent} with the Hall law, i.e., such that
$\pi_i=(p_i)_0-(1/B)\varepsilon_{ij}E_j(x_0,t_0)\neq0$.
No Hall motion can start with such initial conditions;
the consistency of the equations of motion (i.e., $(\dot\vp,\dot\vx,\dot{t})$
lies in the $1$-dimensional
kernel of the singular two-form $\omega-dh\wedge{}dt$) can, however,
be maintained requiring $t=t_0=\const$.
Thus, from each such ``forbidden'' point starts a
strange, ``instantaneous motion''.
 For a constant electric field, for example, these
``instantaneous motions'' are circles with radius
$\vert\vec{\pi}\vert/B_c$, centered at
 $(x_i)_0+(1/B_{c})\varepsilon_{ij}\pi_j$.
The physical interpretation of these ``motions'' is still
unclear to us; our conjecture is that they could be related to
the edge motions in the FQHE \cite{QHE}.

While this paper was being completed, there appeared an
article~\cite{GJPP} discussing rather similar issues.
Let us briefly indicate the relation to our work. In \cite{GJPP} the authors
start, following \cite{DJT}, with an ordinary charged particle in a planar
magnetic field, and then set the mass to zero. Their model becomes
non-commutative
only after taking the limit $m\to0$, and their magnetic field is arbitrary.
Here,
we start with non-commuting the coordinates,
and fine-tune the magnetic field to yield vanishing effective
mass, $m^*=0$. The observation in \cite{GJPP} saying that some
functions have vanishing Poisson bracket with the
dynamical variables is consistent with our two-form
(\ref{symplecticmatrix}) becoming singular.
Then Guralnik et al. present a
non-commutative magnetohydrodynamical model, analogous to ours in the
second paper in \cite{DH}.
Their constraint $\vec{\pi}=0$ corresponds to our lowest Landau level
condition (\ref{LLLcond}).

The recent paper \cite{BNS} contains also some
similar results; they use
another system of canonical coordinates.

\kikezd{\bf Acknowledgement.}
We are indebted to Prof. R.~Jackiw for
correspondence and to Dr. L. Martina for discussions.

%%%%%%%%%%%%%%%%%%%%%%%%%%%%%%%%%%%%%%%%%%%%%%%%%%%%%%%%%%%%%%%%%%%%%%%%%%%%%%
%%%%%%%%%%%%%%%%%%%%%%%%%%%%%%%%%%%%%%%%%%%%%%%%%%%%%%%%%%%%%%%%%%%%%%%%%%%%%%

\end{document}